\documentclass[aps,12pt,nofootinbib,superscriptaddress]{revtex4-1}
\usepackage[latin1]{inputenc}
\usepackage[english]{babel}
\usepackage{graphicx}
\usepackage{color}
\usepackage{amsmath}
\usepackage{amssymb}
\usepackage{multirow}
\usepackage{color}
\graphicspath{{./figs/}}
\begin{document}

\title{Observables of spheroidal magnetized Strange Stars}

\author{D. Alvear Terrero}
\email{dianaalvear@icimaf.cu}
\affiliation{Instituto de Cibern\'{e}tica, Matem\'{a}tica y F\'{\i}sica, \\
 Calle E esq a 15, Vedado 10400, La Habana Cuba}

\author{S. L\'opez P\'erez}
\email{slopez@estudiantes.fisica.cu}
\affiliation{Facultad de F{\'i}sica, Universidad de la Habana,\\ San L{\'a}zaro y L, Vedado, La Habana 10400, Cuba}

\author{D. Manreza Paret}
\email{dmanreza@fisica.uh.cu}
\affiliation{Facultad de F{\'i}sica, Universidad de la Habana,\\ San L{\'a}zaro y L, Vedado, La Habana 10400, Cuba}

\author{A. P\'erez Mart\'{\i}nez}
\email{aurora@icimaf.cu}
\affiliation{Instituto de Cibern\'{e}tica, Matem\'{a}tica y F\'{\i}sica, \\
	Calle E esq a 15, Vedado 10400, La Habana Cuba}

\author{G. Quintero Angulo}
\email{gquintero@fisica.uh.cu}
\affiliation{Facultad de F{\'i}sica, Universidad de la Habana,\\ San L{\'a}zaro y L, Vedado, La Habana 10400, Cuba}

\date{\today}

\begin{abstract}
We study stable spheroidal configurations  of magnetized Strange Stars using an axially symmetric metric in spherical coordinates  that uses a gamma parameter to link the anisotropy in the Equation of State due to the magnetic field with the deformation of the star. The stars are composed by magnetized Strange Quark Matter described within the framework of the MIT-Bag model. Their masses, radii, eccentricity, redshift and mass quadrupole moment are computed. Results are compared with spherical Strange Stars solutions obtained with TOV equations and observational data of Strange Stars candidates. In the spheroidal model the observables depend directly on the deformation of the stars, and even though it is small, the observables strongly deviate from the corresponding spherical configurations. Thus, the highest values of the mass quadrupole moment correspond to the intermediate mass regime. These differences might allow to discriminate between models with/without magnetic field when compared with observations.
\end{abstract}

\maketitle

\section{Introduction}

The core of Neutron Stars (NSs) reaches densities beyond that of nuclear saturation, allowing a phase transition from hadronic to Strange Quark Matter (SQM), which is speculated to be the truly ground state of strong-interacting matter (Bodmer-Witten's conjecture)~\cite{Bodmer, Witten}. Experimentally, the quark--gluon plasma, in the limit of high temperatures and low densities, has been explored in the Relativistic Heavy Ion Collider (RHIC) at Brookhaven National Laboratory (BNL) and in the Large Hadron Collider (LHC) at CERN. Meanwhile, the opposite regime of low temperatures and high densities will be studied in two experimental facilities that are currently under construction, the Facility for Antiproton and Ion Research (FAIR) at GSI and the Nuclotron-based Ion Collider Facility (NICA) at JINR. Nevertheless, since NSs cores might be the natural habitat for SQM in the latter regime, the properties of matter under those conditions could also be inferred through astronomical observations.

A subclass of Neutron Stars composed by a gas of degenerated quarks was first proposed by Itoh in $1970$ \cite{Itoh}, and since then many microscopic models have been proposed for those objects.  The main difference between models are how the strong interaction is described and  what type of SQM phases is considered inside the star \cite{Camenzind}.
Compact objects with quark matter could then be either SSs composed entirely by SQM, or hybrid stars when there is a core of quark matter surrounded by a layer of hadrons.
SSs models are interesting because they describe compact objects with maximum masses around $1.5 \, M_{\odot}$ and radius of $4-8$~km that could account for some observations of cold, dense and small Compact Objects (COs) that do not fit the standard NSs models (see Table \ref{candidatas}) \cite{Lattimer2003, Abubekerov, Taparati, Guver2010_1, Guver2010_2, Elebert, Rawls}. On the other hand, Hybrid Stars have received a lot of attention  because they can reach $2 M_{\odot}$, which is a robust observational constrain on NSs mass \cite{Demorest}.  Recently, these models have gained a new support due to the works by Eemeli Annala et al.\cite{Annala}, where a wide set of theoretical EoS from particle and nuclear physics is compared with benchmark results stemming from Gravitational Waves (GWs) measurements of NSs collisions. According to their results, a star with mass $\sim 2 \, M_{\odot}$ and radius $\sim 12$~km is more likely to have a quark core of approximately $6.5$~km, than to be formed exclusively by baryons.
\begin{table*}[ht!]
	\caption{Masses ($M$) and  radii ($R$) for observed candidates of SSs.}
   	\label{candidatas}
	\centering
	\begin{tabular}{|l|l|l|}
		\hline
		Observed Object & $M(M_{\odot})$  & R(km)\\
		\hline
		\hline
		RXJ185635-3754 \cite{Lattimer2003} & $1.20\pm0.2$ & $8.00\pm1$  \\
   		HerX-1 \cite{Abubekerov, Taparati} & $0.85\pm0.15$ & $8.1\pm0.4$ \\
   		4U1608-52 \cite{Guver2010_1} & $1.74\pm0.14$ & $9.3\pm1$ \\
   		4U1820-30 \cite{Guver2010_2} & $1.58\pm0.06$ & $9.1\pm0.4$ \\
   		SAXJ1808.4-3658 \cite{Elebert} & $0.9\pm0.3$ & $7.9\pm1$ \\
   		Vela X-1 \cite{Rawls} & $1.77\pm0.08$ & $9.56\pm0.08$ \\
   		4U1538-52 \cite{Rawls} & $0.87\pm0.07$ & $7.87\pm0.21$ \\
   		SMC X-1 \cite{Rawls} & $1.04\pm0.09$ & $8.3\pm0.2$ \\
   		LMC X-4 \cite{Rawls} & $1.29\pm0.05$ & $8.83\pm0.09$ \\
   		Cen X-3 \cite{Rawls} & $1.49\pm0.08$ & $9.18\pm0.13$ \\
   \hline
	\end{tabular}
\end{table*}

The study of SSs starts by seeking the SQM Equation of State (EoS). Inside SSs quarks might be unpaired or paired forming a color superconductor state \cite{Ferrer}. However, so far there are neither \textit{ab initio} nor perturbative QCD calculations that leads to the desired EoS in the regime of high densities and zero temperature\cite{Buballa:2003qv}. Consequently, SQM inside COs is usually described through phenomenological models that mimic the main features of QCD \cite{Buballa:2003qv}. One of the most used is the MIT--Bag model \cite{Chodos}, where quarks are considered as quasi-free particles confined into a \textquotedblleft bag\textquotedblright and having fixed masses. This model reproduces confinement and asymptotic freedom with the use of only one external parameter, the bag energy $\mathtt{B_{Bag}}$.

On the other hand, Compact Objects have extreme magnetic fields \cite{Shapiro}. The observed surface magnetic fields in Neutron Stars are in the range of $10^{9}-10^{15}$ G \cite{1993ApJ...408..194T, 1998Natur.393..235K}, while their inner magnetic fields are estimated to be as high as $5\times 10^{18}$ G \cite{Daryel2015, Lattimerprognosis}. As it is well known, the energy momentum tensor of matter under a magnetic field is anisotropic \cite{Ferrer}, a feature that can be interpreted as the gas exerting different pressures in the parallel and perpendicular directions with respect to the magnetic axis \cite{Bali:2014kia}, and that leads to non-spherical stars \cite{Konno:1999zv,Chatterjee:2014qsa,Rizaldy,prd}.
To model the structure of magnetized COs, we have previously used Tolman-Oppenhimer-Volkoff (TOV) equations independently for each pressure \cite{Felipe2008, Felipe2009JPhG,ARD,Perez2019}, and an approximate solution of Einstein's equations for an axially symmetric metric in cylindrical coordinates\cite{Daryel2015}. This leads to the description of two different stellar sequences (one for each pressure), which prevents  to calculate the total mass of the star. That is why, as an attempt to properly describe the macroscopic structure of magnetized COs,  we derived a set of TOV-like structure equations from an axially symmetric metric in spherical coordinates, the $\gamma$-equations \cite{prd}, that allow to describe spheroidal objects as long as their shape is nearly spherical.  These equations  have been used to model White Dwarfs \cite{prd}, the hypothetical magnetized BEC stars~\cite{Gretel} and, preliminary, the Strange Stars \cite{Perez2019}.

In this paper we returned to SSs \cite{Perez2019,Felipe2009JPhG,Felipe2008}, emphasizing in the magnetic field effects on the stability of the SQM and the spheroidal stellar configurations. In addition, we compute the eccentricity, the redshift and the mass quadrupole moment and compare them with those of spherical Strange Stars. This study is essential and is also the starting point of a more ambitious project related to the study of magnetized Hybrid Stars.

In Section \ref{sec2} we analyze the stability of magnetized SQM in astrophysical conditions and its dependence on the density, the bag energy and the magnetic field. Besides, we revisited the magnetized  EoS for SSs and discuss the magnetic field effects on the energy density and pressures. The $\gamma$-equations are presented in Section \ref{sec3} with their corresponding numerical results for magnetized Strange Stars. Section \ref{sec4} is devoted to the computation of the redshift and the mass quadrupole moment. Concluding remarks are given in Section \ref{sec5}.

\section{EoS of magnetized Strange Stars}\label{sec2}

We consider a Strange Star composed by SQM and electrons under the action of a uniform and constant magnetic field oriented in the $z$ direction ${\bf B} = (0,0,B)$. As pointed out before, we use the
 phenomenological MIT--Bag model \cite{Chodos}, where quarks are assumed as quasi-free particles confined into a \textquotedblleft bag\textquotedblright\, that reproduces the asymptotic freedom and confinement through the B$_{\text{bag}}$ parameter, a binding energy which is added to the quarks energy and subtracted from their pressure \cite{Chodos}.

For a magnetized gas of quarks and electrons, the pressure and the energy density are obtained from the thermodynamical potential \cite{Felipe2008}
\begin{equation}\label{Thermo-Potential}
\Omega_f(B,\mu_f,T)  = -e_fd_f B T\int_{-\infty}^{\infty} \! \frac{dp_3}{4\pi^2}  \sum_{l=0}^{\infty}g_l
  \times\sum_{p_4}\ln\bigg[(p_4+i\mu_f)^2+ \varepsilon^2_{lf} \bigg],
\end{equation}
where $l$ stands for the Landau levels and $f=e, u, d, s$ for the electrons and each quark flavour; $d_f$ is the flavour degeneracy factor\footnote{$d_e=1$ and $d_{u,d,s}=3$.} and $g_l=2-\delta_{l0}$ includes the spin degeneracy of the fermions for $l\neq0$. Moreover, $T$ is the absolute temperature, while $\mu_f$, $m_f$ and $e_f$  are the chemical potential, the mass and the charge of each particle respectively. The spectrum of charged fermions coupled to a magnetic field is $\varepsilon_{lf}=\sqrt{p_3^2+2|e_fB|l+m_f^2}$ \cite{Felipe2008}.

Eq.~\eqref{Thermo-Potential} can be divided in two contributions
\begin{equation}\label{Thermo-Potential-2}
\Omega_f (B,\mu_f,T)=\Omega^\text{vac}_f(B,0,0)+\Omega^\text{st}_f(B,\mu_f,T),
\end{equation}
with
\begin{eqnarray}
\Omega^\text{vac}_f(B,0,0) &=&-\frac{e_fd_f B}{2\pi^2}\int_{0}^{\infty}\hspace{-2mm}dp_3\sum_{l=0}^{\infty}g_l\varepsilon_{lf} \label{vaccon},\\\label{omegast}
\Omega^\text{st}_f(B,\mu_f,T) &=& -\frac{e_fd_f BT}{2\pi^2}\int_{0}^{\infty}\hspace{-2mm}dp_3\sum_{l=0}^{\infty}g_l 
\ln\hspace{-1mm}\big[1+e^{-\beta(\varepsilon_{lf}\pm\mu_f)}\big]
\label{stcon}\hspace{-0.5mm}.
\end{eqnarray}
 The vacuum contribution Eq.~\eqref{vaccon} does not depend on the chemical potential nor on the temperature and presents an ultraviolet divergence that must be renormalized \cite{Schwinger}, resulting in
\begin{equation}\label{vaccon2}
\Omega^\text{vac}_f(B,0,0)= \frac{d_fm_f^4}{24\pi^2}\left(\frac{B}{B^c_f}\right)^2\text{ln}\frac{B}{B^c_f}.
\end{equation}
In Eq.\eqref{vaccon2} $B^c_f=m^2_f/e_f$ is the  critical magnetic field\footnote{The magnetic field at which the cyclotron energy of the particles is comparable to their rest mass. For electrons this is the so-called Schwinger field.}. For electrons $B^c_e\sim10^{13}$~G while for quarks up, down and strange we have $B^c_u\sim10^{15}$ G, $B^c_d\sim10^{16}$ G and $B^c_s\sim10^{19}$ G, respectively.

COs have temperatures much smaller than the Fermi temperature of the gases that compose them \cite{Camenzind}. Hence, a good approximation is to compute the thermodynamical potential of these gases in the degenerate limit (T$\rightarrow0$) \cite{Shapiro, Camenzind}. In that case, the statistical part of the thermodynamical potential becomes
\begin{equation}\label{stcon2}
\Omega^\text{st}_f(B,\mu_f,0)=-\frac{e_fd_f B}{2\pi^2}\int_{0}^{\infty}dp_3\sum_{l=0}^{\infty}g_l\Theta(\mu_f-\varepsilon_{lf}),
\end{equation}
where $\Theta(\zeta)$ is the unit step function. From  Eq.~\eqref{stcon2}, we obtain
\begin{equation}\label{stcon3}
  \Omega^\text{st}_f(B,\mu_f,0)=-\frac{e_fd_f B}{4\pi^2}\sum_{0}^{l_{max}}g_l\bigg[\mu_fp_f^l
  -(\varepsilon_f^l)^2\text{ln}\bigg(\frac{\mu_f+p_f^l}{\varepsilon_f^l}\bigg)\bigg],
\end{equation}
where $p_f^l=\sqrt{\mu_f^2-(\varepsilon_f^l)^2}$ is the fermi momentum of the particles, $\varepsilon_f^l=\sqrt{m_f^2+2qBl}$ their ground state energies and $l_{\text{max}}=I[\frac{\mu_f^2-m_f^2}{2e_fB}]$ the maximum number of occupied Landau levels for fixed magnetic field and chemical potentials. $I[z]$ denotes the integer part of $z$.

Due to the high fermionic densities in the stars interior, the vacuum contribution Eq.~\eqref{vaccon2} is negligible with respect to the statistical one,  Eq.~\eqref{stcon3}~\cite{Ferrer}.
Therefore, the thermodynamical potential of the degenerate fermion system can be approximated to $\Omega_f(B,\mu_f,0)=\Omega^\text{st}_f(B,\mu_f,0)$.

Strange Quark Matter inside the star must be in stellar equilibrium \cite{Felipe2008}. So, we impose $\beta$ equilibrium, charge neutrality and baryon number conservation  to the system in terms of the particle densities $N_f=-\partial\Omega_f/\partial\mu_f$ and the chemical potentials. These conditions read \cite{Felipe2008}
\begin{subequations}\label{eqsqm}
    \begin{eqnarray}
    \mu_u+\mu_e-\mu_d=0 \,\,\,, \,\,\, \mu_d-\mu_s&=& 0,\\
    \sum_f  e_fN_f=0 \,\,\,, \,\,\
    \sum_{i=u,d,s}N_i=3N_B.
    \end{eqnarray}
\end{subequations}

With these considerations, the magnetized SSs EoS are
\begin{subequations}\label{EoSSS}
    \begin{align}
    E &= \sum_{f}\left[\Omega_f+ \mu_fN_f\right]+B_{\text{bag}} +\frac{B^2}{8\pi},\label{EoS1}\\
    P_{\parallel} &=- \sum_{f}\Omega_f-B_{\text{bag}}-\frac{B^2}{8\pi}, \label{EoS2}\\
    P_{\perp} &= -\sum_{f}\left[\Omega_f+B\mathcal{M}_f\right]-B_{\text{bag}}+\frac{B^2}{8\pi},\label{EoS3}
    \end{align}
\end{subequations}
where $\mathcal{M}_f=-\partial\Omega_f/\partial B$ is the magnetization. In Eqs.~\eqref{EoSSS} we can easily identify three different contributions. The first one is given by the sum over the thermodynamical quantities of the species and corresponds to the statistical contribution of each kind of particle. In the case of the perpendicular pressure Eq.~\eqref{EoS3}, this term includes a contribution that comes from the particle magnetization: $-B\mathcal{M}_f$ \cite{Chaichian}. The second terms in the EoS, $\pm B_{\text{bag}}$, are the ones ensuring asymptotic freedom and confinement for quarks \cite{Witten,Buballa:2003qv}. Finally, the last terms in Eqs.~\eqref{EoSSS} are the magnetic field pressures and energy density $P^B_{\perp}=E^B=-P^B_{\parallel}=B^2/8\pi$ \cite{Ferrer}, or Maxwell contribution. These terms are included since they also participate in the gravitational stability of the star.

Eqs.~\eqref{EoSSS} allow to analyze the influence of the magnetic field in the stability of SQM, so that
\begin{equation}\label{des}
\left.\frac{E}{n_B}\right|_{SQM}^B<\left.\frac{E}{n_B}\right|_{SQM}^{B=0}<\left.\frac{E}{n_B}\right|_{\mathrm{Fe}^{56}} = 930~\text{MeV/fm}^3,
\end{equation}
is fulfilled inside the star. Fixing $\left.E/n_B\right|_{SQM}^B=930$~MeV/fm$^3$ lead us to the stability window of SQM in the plane $\mathbf{B}$~vs~$\mathtt{B_{Bag}}$.
\begin{figure}[h]
	\centering
	\begin{tabular}{cc}
		\includegraphics[width=0.48\textwidth]{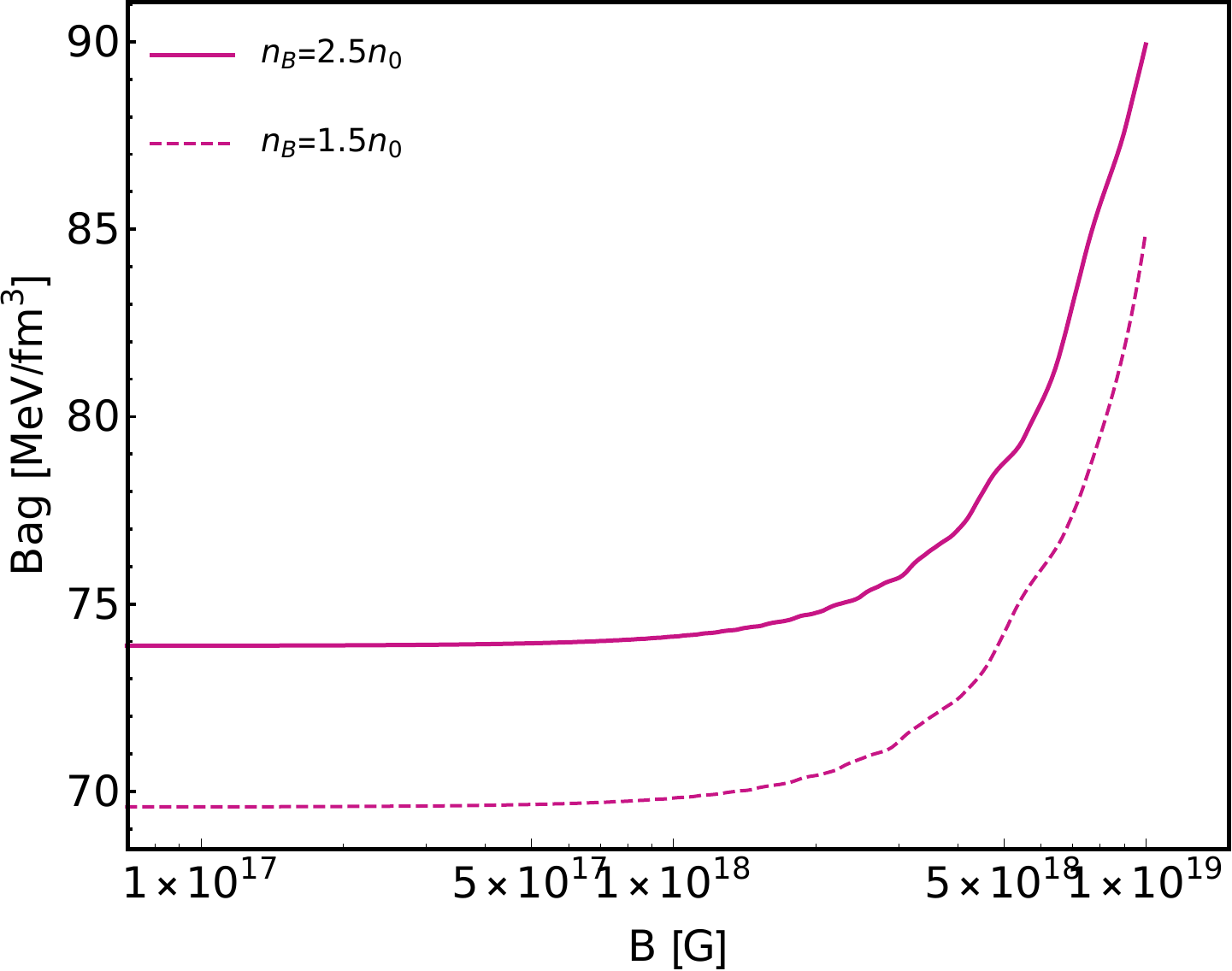}
		&  \includegraphics[width=0.51\textwidth]{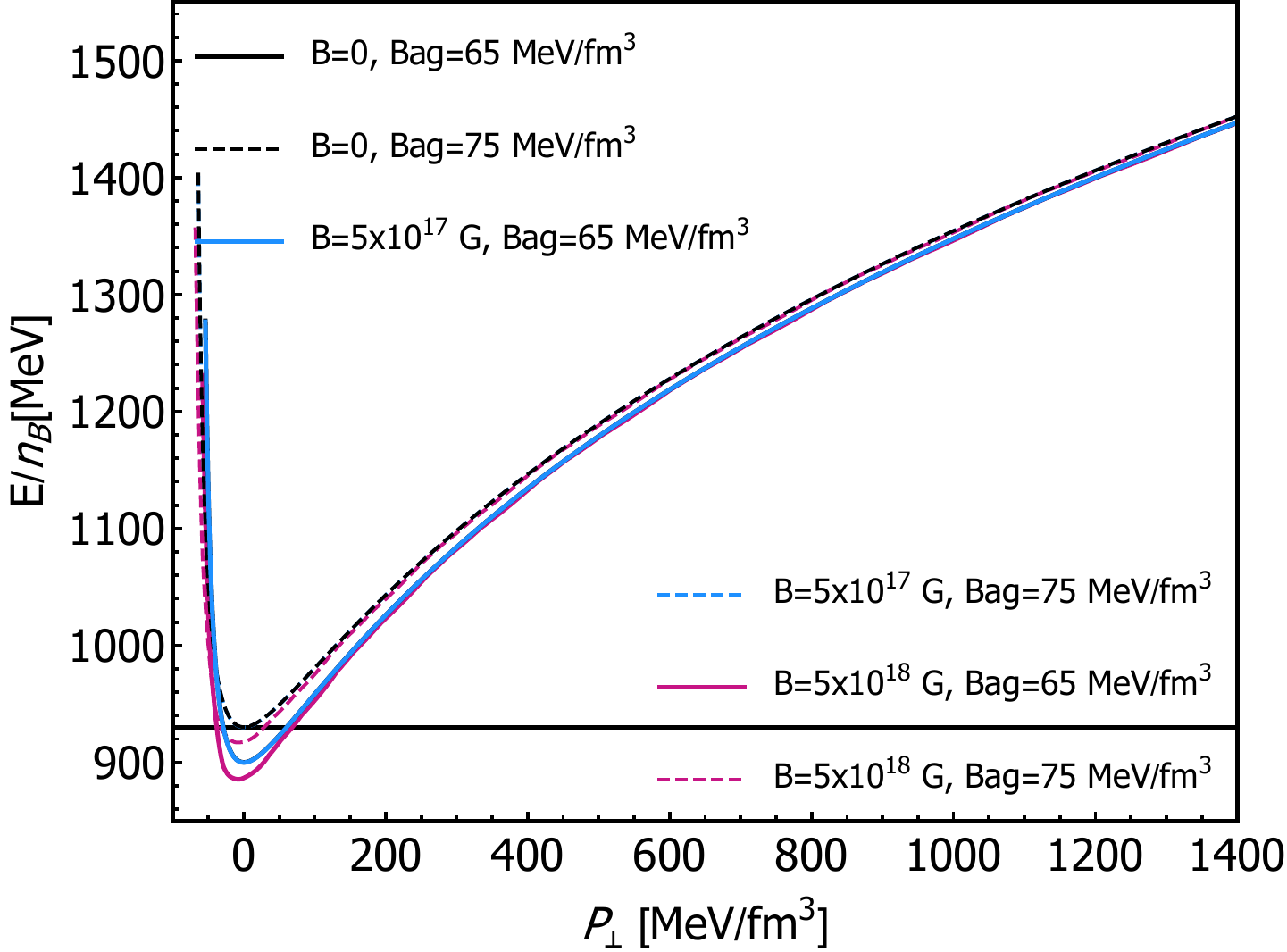}
	\end{tabular}
	\caption{Left panel: Stability window for magnetized SQM in the plane $\mathbf{B}$ vs $\mathtt{B_{Bag}}$. Right panel: Energy per baryon ($E/n_B$) as a function of the perpendicular pressure ($P_{\perp}$) at $B=[0, 5\times10^{17}, 10^{18}]$~G and for fixed values of the  $\mathtt{B_{Bag}}=[65, 75]$~MeV/fm$^3$.}
	\label{bagdepend}
\end{figure}
This is shown in the left panel of Fig. \ref{bagdepend}, where the SQM is stable for the pairs of $\mathbf{B}$~vs~$\mathtt{B_{Bag}}$ below the curves and unstable otherwise. We can also see from this plot that increasing the magnetic field and the baryon density augments the stability region. So, we can say that magnetic field  contributes to stabilize SQM with respect to nuclear matter in astrophysical environments, favoring the existence of SSs.
This conclusion is reinforced by the results shown in
the right panel of Fig. \ref{bagdepend}, that illustrates the energy per baryon as a function of the perpendicular pressure. At $P_{\perp}=0$, $\left.E/n_B\right|_{SQM}^B<\left.E/n_B\right|_{\mathrm{Fe}^{56}}$, i.e. the magnetic field increases the stability region, while the $\mathtt{B_{Bag}}$ diminishes it.

A more detailed stability analysis should take into account the $\mathtt{B_{Bag}}$ dependency with the magnetic field and the baryon density \cite{Ferrer}. Nevertheless, the theoretical attempts to find those dependencies or to constrain the $\mathtt{B_{Bag}}$ values are yet inconclusive and very model dependent~\cite{Buballa:2003qv}. Therefore, in what follows we take $\mathtt{B_{Bag}}$ as an independent external parameter and study SSs for two fixed, and reasonable, values of it: $65$~MeV/fm$^3$ and $75$~MeV/fm$^3$.

\begin{figure}[!h]
\begin{tabular}{cc}
    \includegraphics[width=0.48\textwidth]{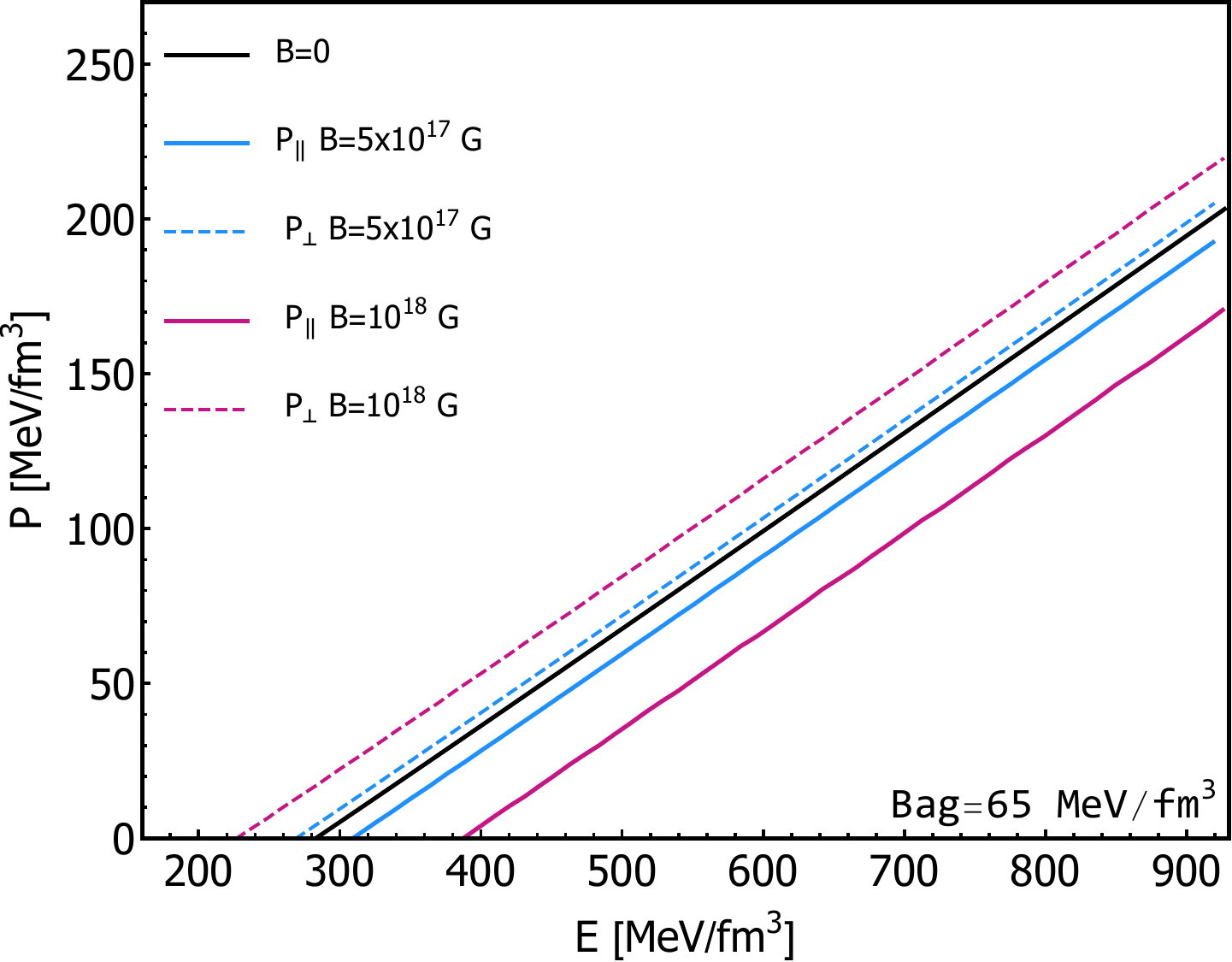}
    &\includegraphics[width=0.48\textwidth]{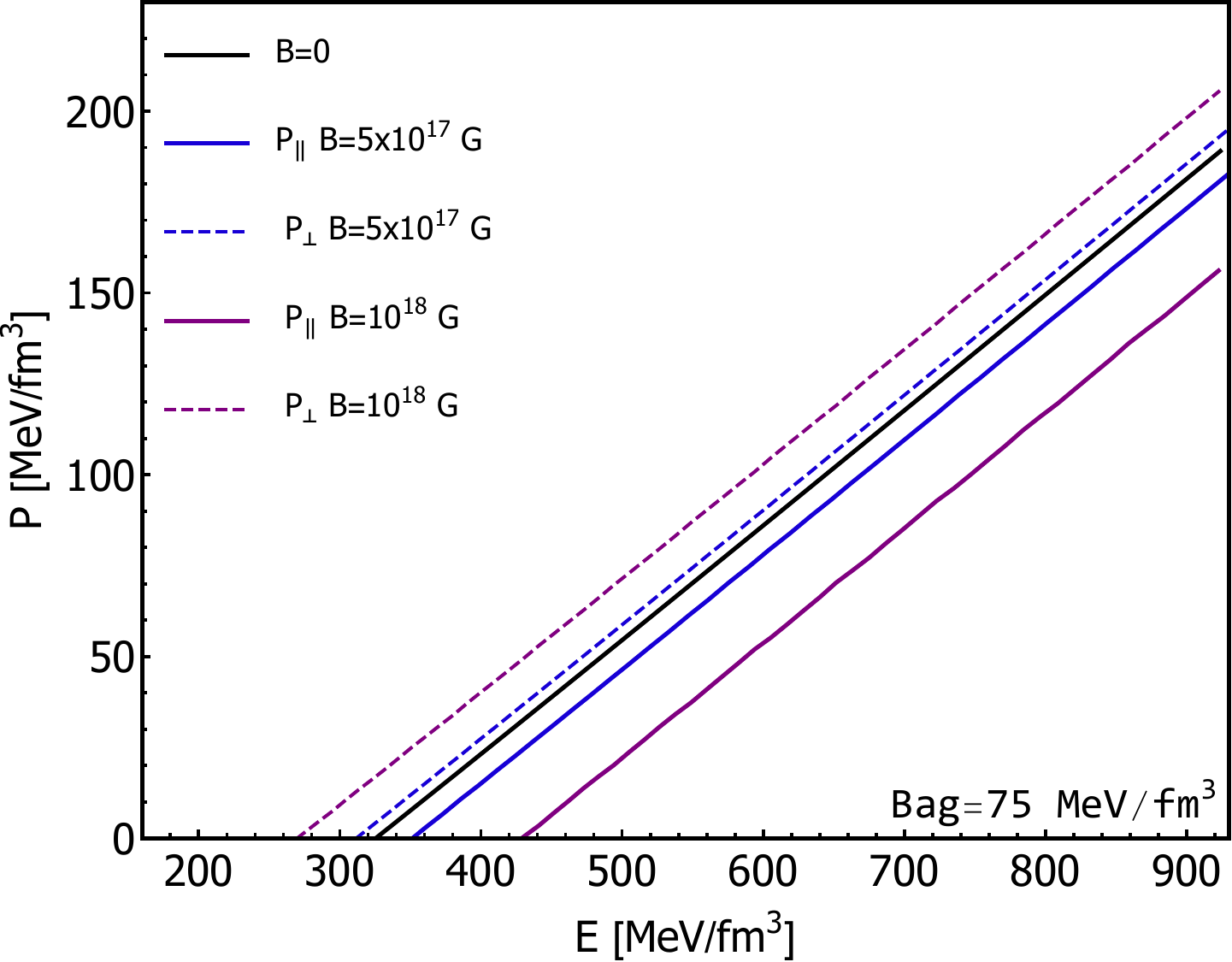}
    \end{tabular}
    \caption{EoS for magnetized SSs at $B=[0, 5\times10^{17}, 10^{18}$~G] and for fixed values of the  $\mathtt{B_{Bag}}=[65$~MeV/fm$^3$, $75$~MeV/fm$^3$].}
    \label{eosplot}
\end{figure}

The SQM EoS obtained for those values of $\mathtt{B_{Bag}}$ at $B=[0, 5\times10^{17},10^{18}]$~G are depicted in Fig. \ref{eosplot}. Note that at higher values of the magnetic field, the difference between the perpendicular and parallel pressures is more appreciable. On the other hand, for a fixed $B$, the pressures decrease when increasing the $\mathtt{B_{Bag}}$.
These effects on the EoS will be reflected in the macroscopic structure of the star as we will see in the next section.
\section{Magnetized Strange Stars: Mass and Radii}\label{sec3}


The axial symmetry imposed in the star by the magnetic field is irreconcilable with the spherical symmetry of standard TOV equations. Consequently, it is desirable the use of axisymmetric metrics if one wishes to describe the structure of magnetized COs (see \cite{prd} and references therein). Here, we follow  and use a set of axisymmetric structure equations derived from the so--called $\gamma$ metric \cite{prd} \begin{align}\label{Ec(13)}
ds^2 = & - \left[1-2M(r)/r\right]^{\gamma}dt^2 + \left[1-2M(r)/r\right]^{-\gamma}dr^2
+ r^2\sin\theta d\phi^2 + r^{2}d\theta^2,
\end{align}
where $\gamma = z/r$ accounts for the deformation of the matter source with respect to the spherical shape and parametrizes the polar radius, $z$, in terms of the equatorial one, $r$.

Starting from this metric and considering the anisotropic energy-momentum tensor of magnetized matter, the following structure equations are obtained \cite{prd}
\begin{subequations}\label{gTOV}
    \begin{eqnarray}
    \frac{dM}{dr}&=& 4\pi r^{2}\frac{(E_{\parallel}+E_{\perp})}{2}\gamma, \label{gTOV1}\\
    \frac{dP_{\parallel}}{dz}&=&\frac{1}{\gamma}\frac{dP_{\parallel}}{dr}\nonumber\\
    &=&-\frac{(E_{\parallel}+P_{\parallel})[\frac{r}{2}+4\pi r^{3}P_{\parallel}-\frac{r}{2}(1-\frac{2M}{r})^{\gamma}]}{\gamma r^{2}(1-\frac{2M}{r})^{\gamma}}, \label{gTOV2}\\
    \frac{dP_{\perp}}{dr}&=&-\frac{(E_{\perp}+P_{\perp})[\frac{r}{2}+4\pi r^{3}P_{\perp}-\frac{r}{2}(1-\frac{2M}{r})^{\gamma}]}{r^{2}(1-\frac{2M}{r})^{\gamma}}. \label{gTOV3}
    \end{eqnarray}
\end{subequations}
Eqs.~(\ref{gTOV}) describe the variation of the mass and the pressures with the spatial coordinate $r$ for an anisotropic axially symmetric CO as long as the parameter $\gamma$ is close to one \cite{prd}. Note that they are coupled through the dependence with the energy density and the mass.

To solve Eqs.~(\ref{gTOV}), one starts from a point in the star center with $\mathcal E_c = E(r=0)$, $P_{\parallel_c} = P_\parallel(r=0)$ and $P_{\perp_c} = P_\perp(r=0)$, taken from the EoS, and ends the integration at $P_\parallel (Z) = 0$  and $P_\perp (R) = 0$, the conditions that defines the radii of the star. But before doing so, one should deal with the fact that $\gamma$ is a free parameter that can not be obtained from Eqs.\eqref{gTOV}. To overcome this difficulty in \cite{prd} we proposed the ansatz $\gamma = P_{\parallel_c}/P_{\perp_c}$, that connects the geometry of the system with the anisotropy produced by the magnetic field. This interpretation of $\gamma$ is based on the fact that for spherical stars a lower central pressure leads to a smaller radius for a fixed central energy, together with the relation $\gamma = z/r$. This ansatz implies that the shape of the star is only determined by the anisotropy of the EoS in its center and neglects the dependence of the deformation on the pressures inner profiles. However, its use gives reasonable results as long as $\gamma \simeq 1$  \cite{prd}. In addition, when setting $B=0$, $P_{\perp}=P_{\parallel}$, $\gamma=1$ and the standard non-magnetized solution for the structure of COs (TOV equations) is recovered.

\begin{figure}[h!]
	\centering
	\begin{tabular}{cc}
		\includegraphics[width=0.5\textwidth]{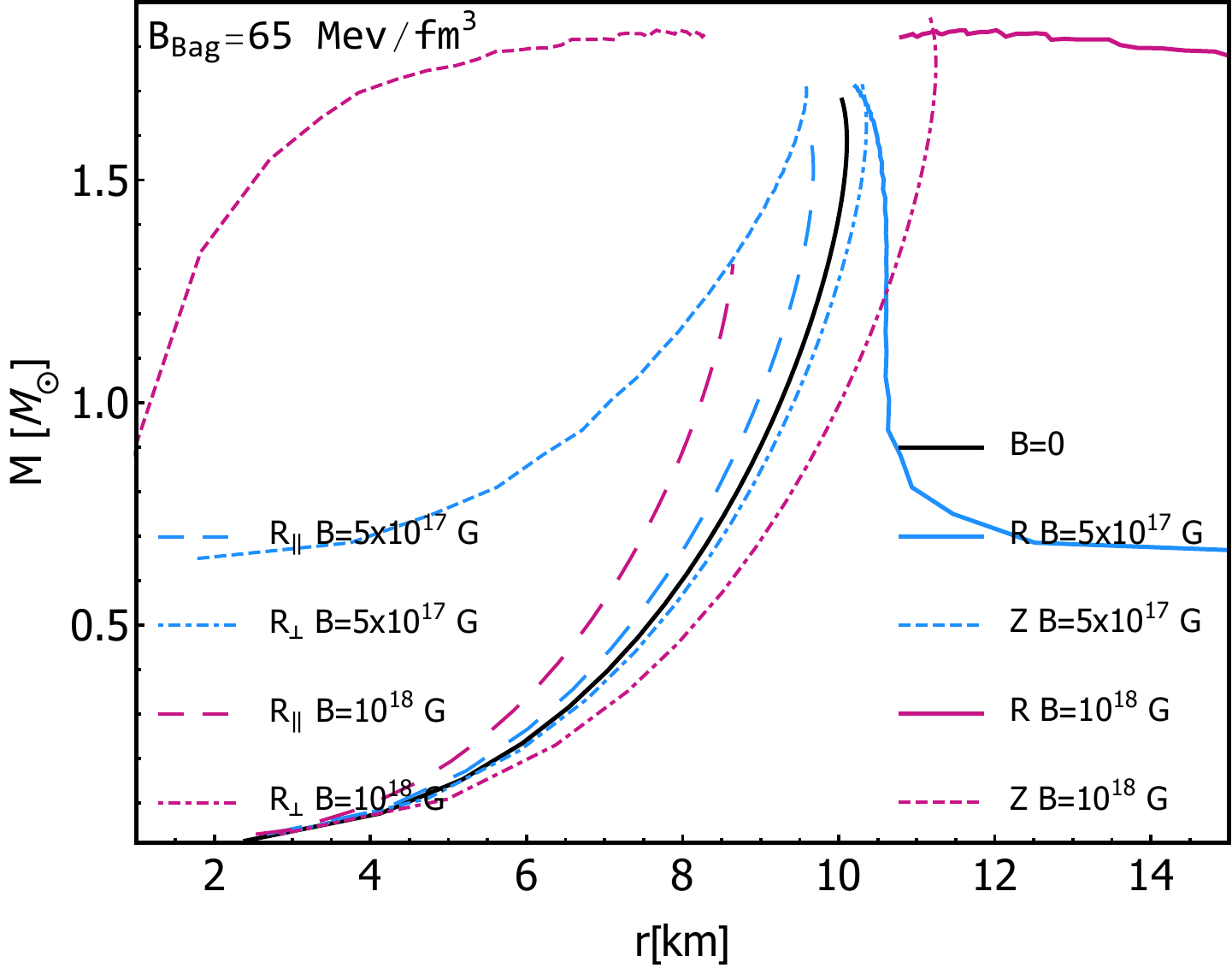}
		& \includegraphics[width=0.5\textwidth]{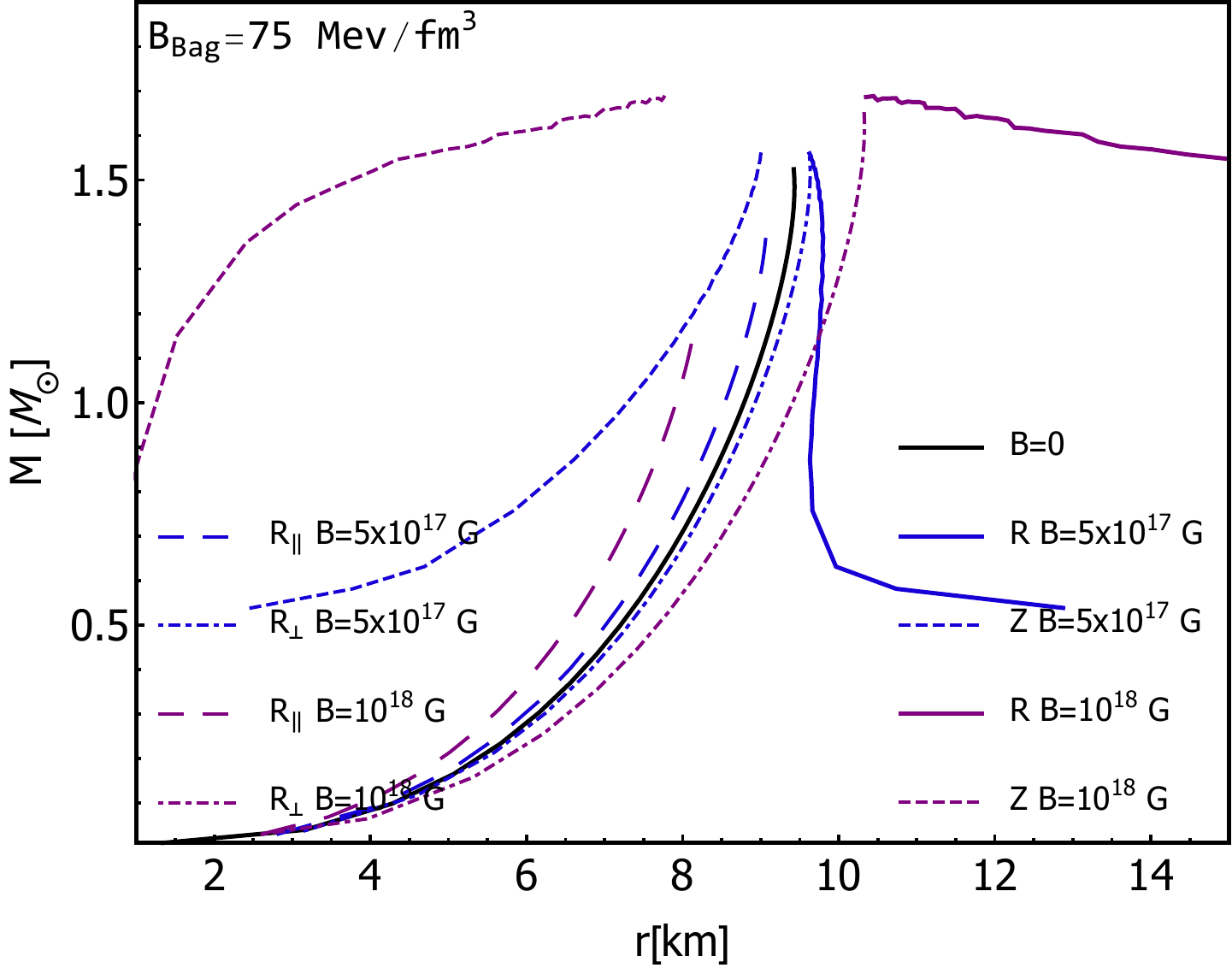}
	\end{tabular}
	\caption{Solutions for spheroidal configurations in comparison with TOV solutions and the non-magnetized configuration at $B=5\times10^{17}$~G and $B=10^{18}$~G, where $r$ represents both radii $R$ and $Z$. Left panel: $\mathtt{B_{bag}}=65$~MeV/fm$^3$. Right panel: $\mathtt{B_{bag}}=75$~MeV/fm$^3$~\cite{Perez2019}.}
	\label{gammaMR}
\end{figure}

In Fig.~\ref{gammaMR} we show the solutions of Eqs.~\eqref{gTOV} with the use of Eqs.~\eqref{EoSSS} for several values of the magnetic field and the bag energy. They are compared with the non-magnetized case and with the TOV solutions considering the pairs ($E,P_{\perp}$) and ($E,P_{\parallel}$) as independent EoS. The maximum masses and corresponding radii are shown in Table \ref{mrmaxg}. In the case of TOV solutions, using one EoS or the other leads to different mass-radius relations, whose differences increase with the magnetic field. For a given energy density range, a higher pressure implies bigger and massive stars.  Also, for a fixed value of the magnetic field and the $\mathtt{B_{Bag}}$, the difference in the stars size obtained with the pairs ($E,P_{\perp}$) and ($E,P_{\parallel}$), is larger for heavier stars. This suggest that more massive stars will have a greater deformation.

Unlike TOV equations, Eqs.~\eqref{gTOV} allow us to model the star as a spheroidal with an equatorial radius $R$ and a polar radius $Z$. So, in Fig.~\ref{gammaMR} the $M-R$ and $M-Z$ curves correspond to an unique sequence of  stars, while the $M-R_{\perp}$ and the $M-R_{\parallel}$ ones stand for two different sequences with EoS ($E,P_{\perp}$) and ($E,P_{\parallel}$) respectively.
The stellar configurations obtained with Eq.~\eqref{gTOV} are oblate objects ($R>Z$), as expected since $P_{\perp}>P_{\parallel}$ (see Figs. \ref{eosplot}, \ref {gammaMR}). Besides, on the contrary of what happens with TOV solutions, for which the difference between $R_{\perp}$ and $R_{\parallel}$ increases with the mass, the deformation of our spheroidal stars -the differences between the equatorial and the polar radius- decreases with the mass \cite{Perez2019} (See Fig.~\ref{gammaMR}). Hence, the importance of building a model, as the one we present, that takes into account both pressures simultaneously.
\begin{table*}[ht!]
	\caption{Maximum values of masses ($M$) Eq.~\eqref{gTOV} and the corresponding radius $R$, $Z$)}
	\label{mrmaxg}
	\centering
	\begin{tabular}{|l|l|l|l|l|}
		\hline
		$B$(G) & $\mathtt{B_{Bag}}$(MeV$/$fm$^3$) & $M(M_{\odot})$ & R(km) & Z(km)\\
		\hline
		\hline
		$5\times 10^{17}$& 65 & 1.74 & 9.79 & 9.41 \\	
		& 75 & 1.63 & 9.08 & 8.79 \\	
		$10^{18}$& 65 & 1.84 & 12.01 & 7.66\\	
		& 75 & 1.69 & 10.20 & 7.88 \\	
		\hline
	\end{tabular}
\end{table*}

The atypical behavior of the curves at $B=10^{18}$~G Fig.~\ref{gammaMR} is because, for this value of the magnetic field the $\gamma$ parameter escapes the range established in \cite{prd}, ($1>\gamma>0.8$) for the validity of Eqs.~\eqref{gTOV}. It is interesting that other models of Neutron Stars EoS and structure equations also present difficulties when treating magnetic fields of the order or higher than $B=10^{18} $~G. For example, in \cite{Daryel2015}, where the anisotropy is considered through a metric in cylindrical coordinates, the metric coefficients diverge for $B \geq 1.8\times 10^{18} $~G. Besides, the theoretical limit established by the Virial theorem for the maximum magnetic field that can be supported by a NS is precisely $B\simeq10^{18}$~G~\cite{Khar}. Therefore, our results seem to support this limit, indicating the relevance of going deeper in the understanding of the physical reasons behind it.
\begin{figure}[!h]
	\includegraphics[width=0.48\textwidth]{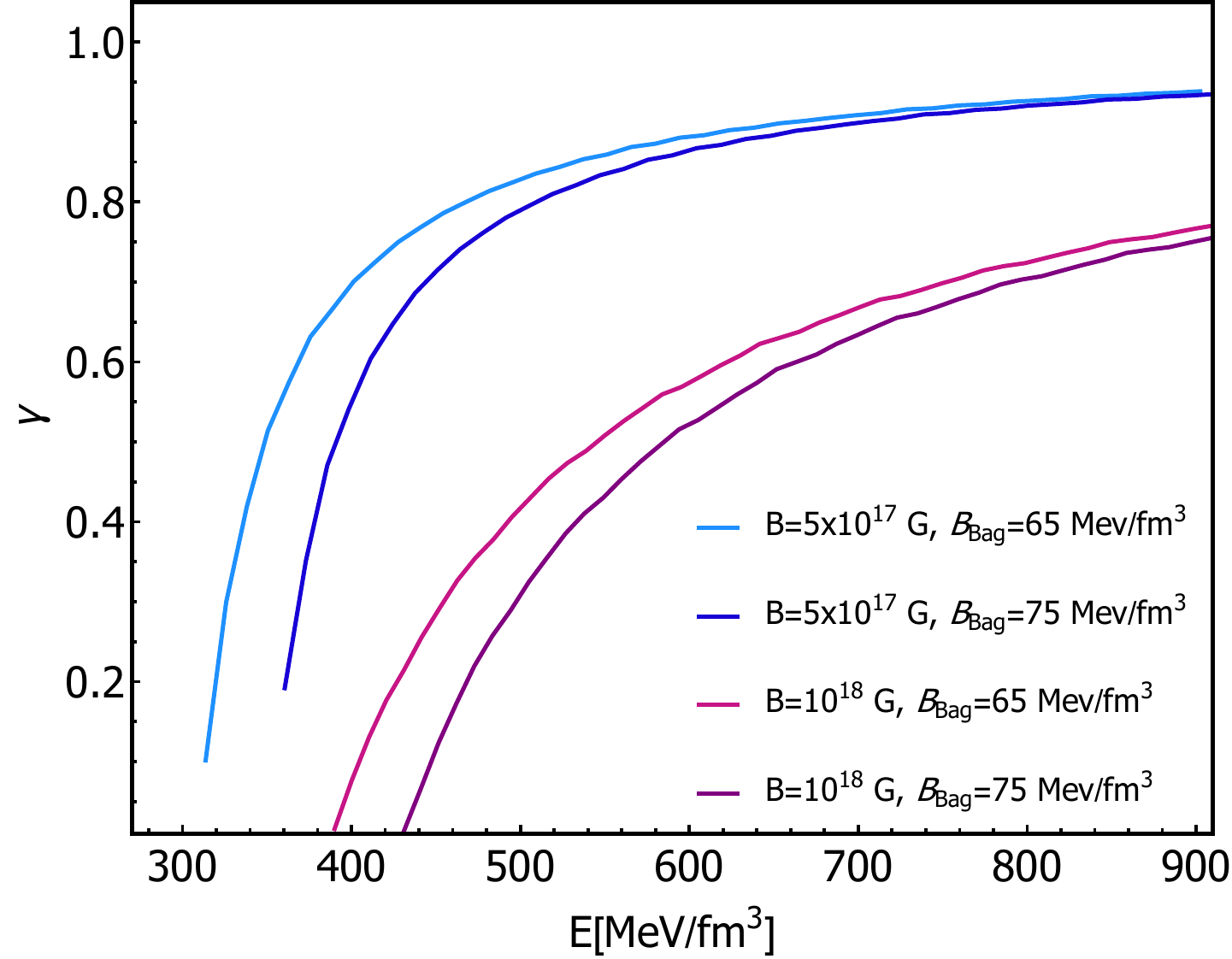}
	\caption{$\gamma$ parameter as a function of central energy density at $B=[0, 5\times10^{17}, 10^{18}$~G] and for fixed values of the  $\mathtt{B_{Bag}}=[65$~MeV/fm$^3$, $75$~MeV/fm$^3$].}    \label{gammaE}
\end{figure}

\begin{figure}[h!]
	\centering
	\begin{tabular}{cc}
		\includegraphics[width=0.5\textwidth]{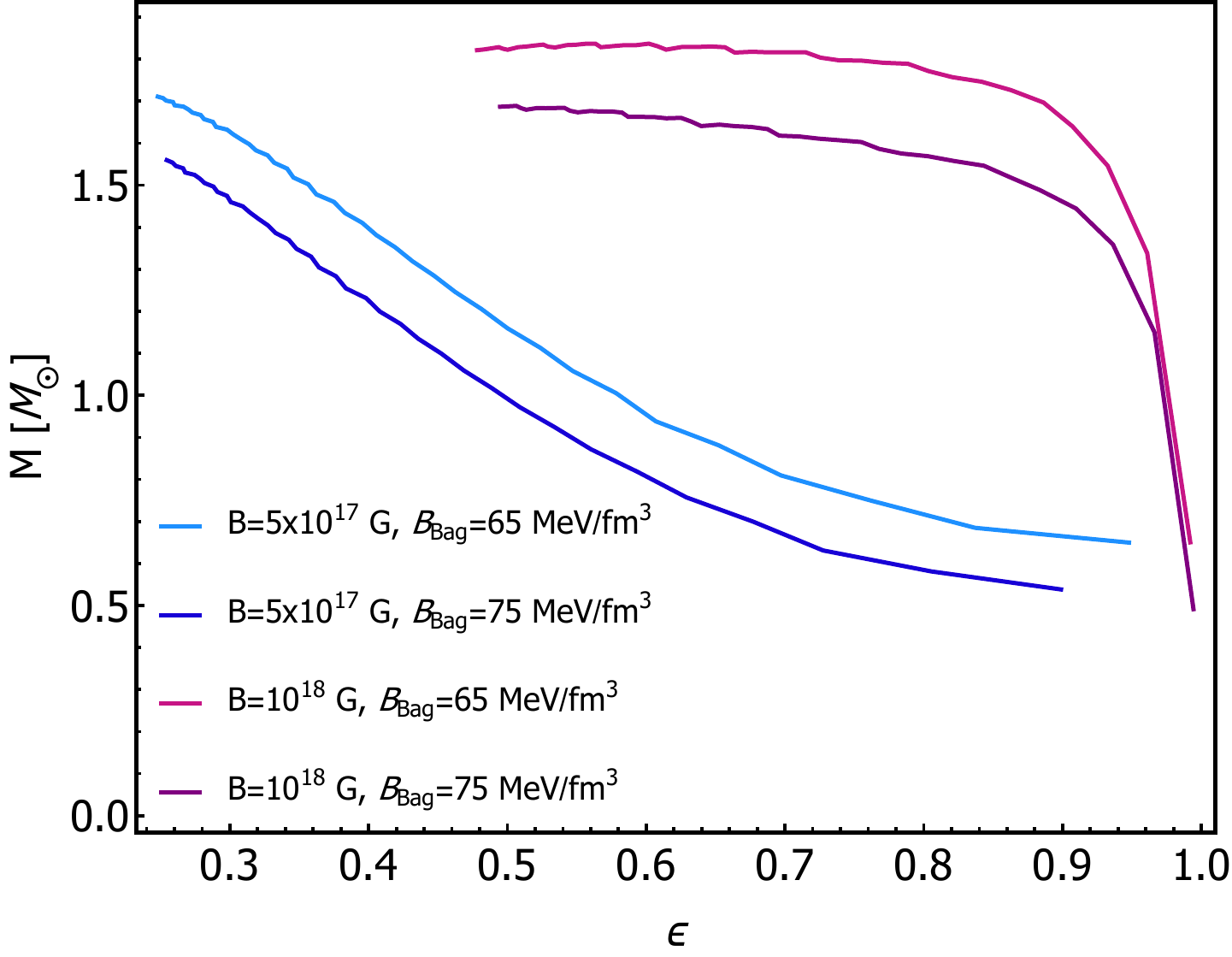} &
	    \includegraphics[width=0.5\textwidth]{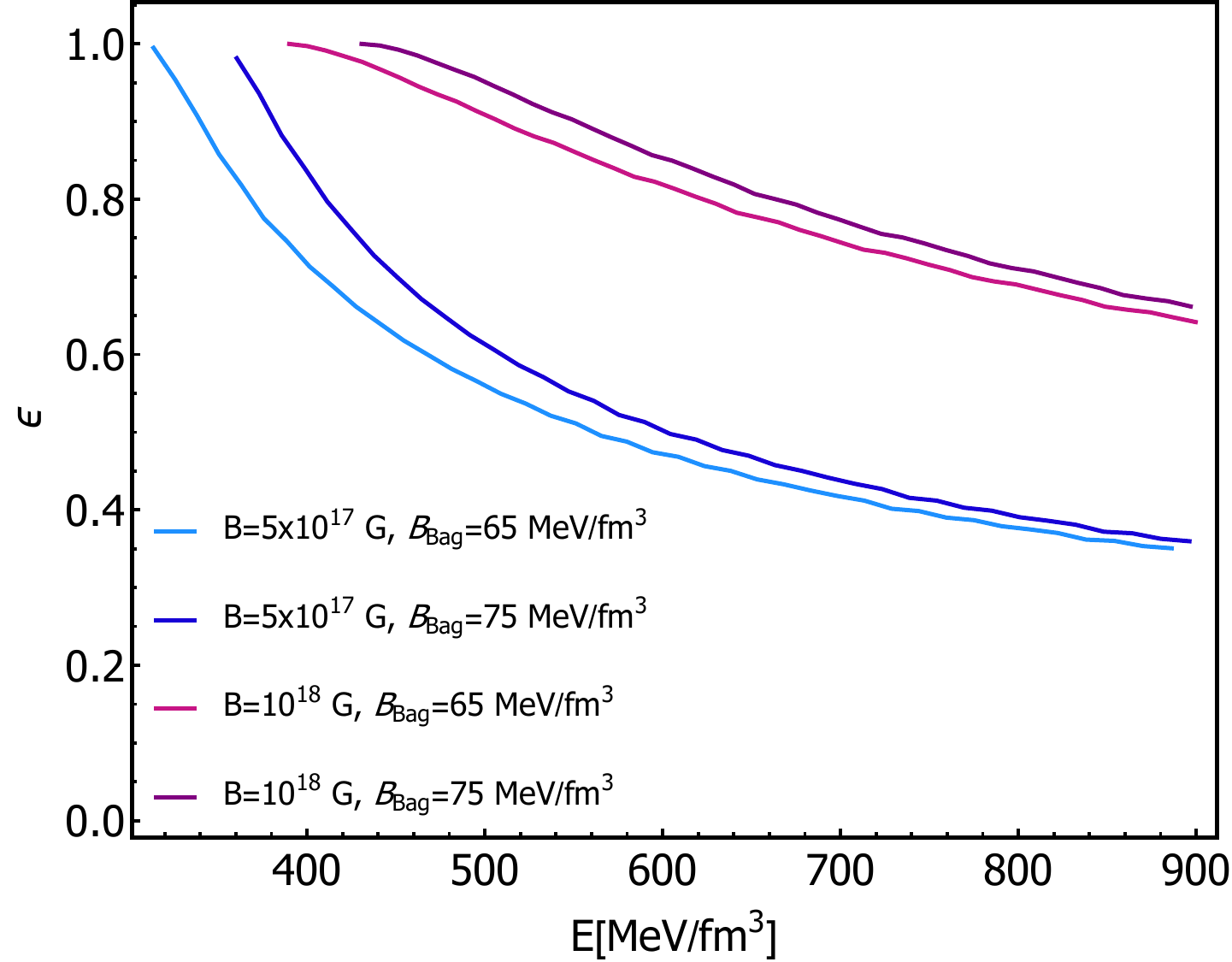}
	\end{tabular}
	\caption{Elipticity ($\epsilon$) as a function of the mass $M$ (left) and the energy density  $E$ (right) at $B=[0, 5\times10^{17}, 10^{18}$~G] and for fixed values of the  $\mathtt{B_{Bag}}=[65$~MeV/fm$^3$, $75$~MeV/fm$^3$].}
	\label{emgamma}
\end{figure}

The effects of varying $\mathtt{B_{Bag}}$ and the magnetic field on the deformation of the stars can also be seen through the ellipticity, defined as \cite{Rizaldy}
\begin{equation}\label{elip}
\epsilon=\sqrt{1-\gamma}.
\end{equation}
In the spherical case  $\gamma\rightarrow 1$ and $\epsilon\rightarrow 0$, while for the most deformed stars, $\gamma\rightarrow 0$ and $\epsilon\rightarrow 1$. In Fig.~\ref{emgamma} we show the ellipticity $\epsilon$ as a function of the  mass and the central energy density, and show that the increase of the magnetic field and the $\mathtt{B_{Bag}}$ leads to a greater deformation of the star.
\subsection{Stability of the solutions of the $\gamma$--equations}

The mass-radii curves obtained for SSs with Eqs.~(\ref{gTOV}) correspond to stellar configurations in hydrostatic equilibrium \cite{Glendenning}. However, equilibrium does not necessarily imply stability~\cite{Shapiro, Gleiser1988}. To find which of those solutions represents stable COs we will use two criteria. The first is related to the stability of the star with respect to radial oscillations and requires that $dM/dE_0>0$ \cite {Shapiro}, so the CO will not fall apart. The second requires the star's gravitational mass (the one calculated with the structure equations) to be lower than its baryonic mass ($M_B$), which is  the sum of the masses of all its particles. This last criterion is applied to guarantee the stability of the CO with respect to the dispersion of the matter that composes it~\cite{Gleiser1988}.

\begin{figure}[h!]
	\centering
	\begin{tabular}{cc}
		\includegraphics[width=0.5\textwidth]{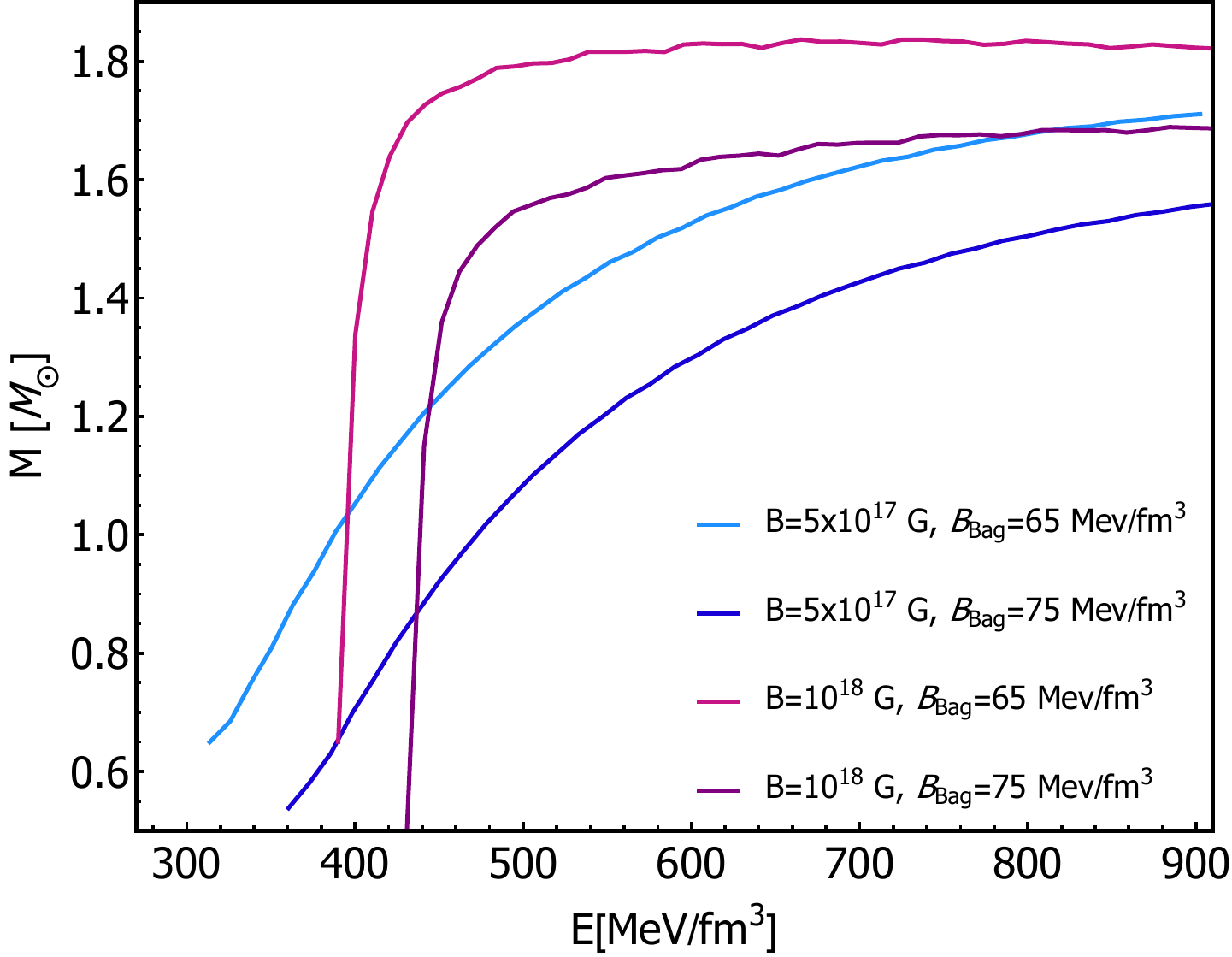}
		& \includegraphics[width=0.5\textwidth]{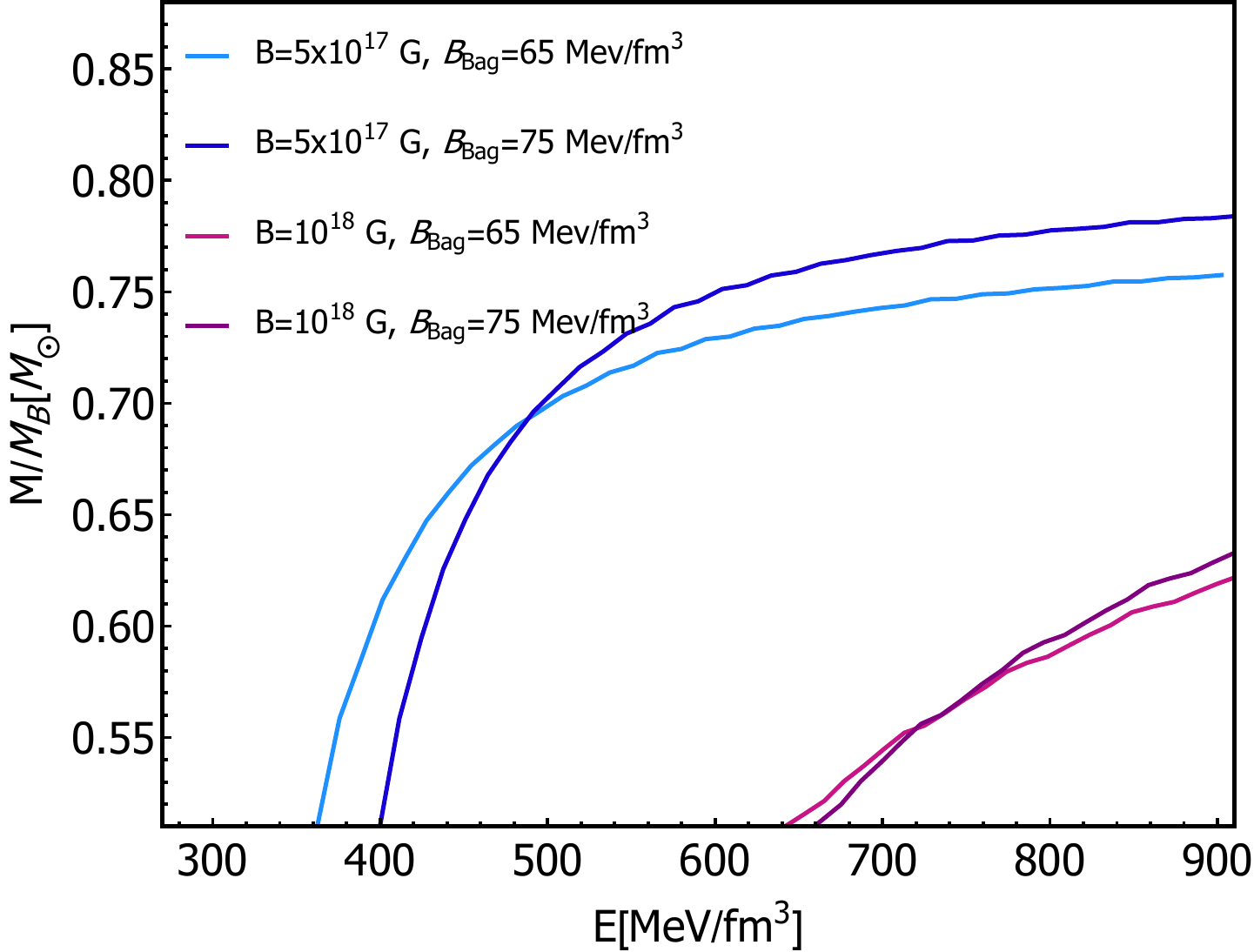}		
	\end{tabular}
	\caption{Mass $M$ as function of energy density  $E$ (left) and ratio between $M$ and the baryon mass $M_B$ (right) at $B=[0, 5\times10^{17}, 10^{18}$~G] and for fixed values of the  $\mathtt{B_{Bag}}=[65$~MeV/fm$^3$, $75$~MeV/fm$^3$].}
	\label{megamma}
\end{figure}

To analyze the stability with respect to radial oscillations of the $\gamma$-equations solutions, in the left panel of Fig.~\ref{megamma} the mass has been plotted as a function of the central energy density. The points represent the maximum of each curve. Therefore, the stars to the left of the maximum are stable ($dM/dE_0>0$), while the ones to the right are not ($dM/dE_0<0$) and must be discarded. For fixed magnetic field, increasing $\mathtt{B_{Bag}}$ increase the stability region, whereas, for a fixed $\mathtt{B_ {bag}}$, increasing $B$ reduces it. Note that the solutions for $B=10^{18}$~G are out of the validity range of $\gamma$--equations for low densities while are unstable with respect to radial oscillations for the higher ones. Therefore, in what follows, we will not consider the observables for this value of the magnetic field.

To apply the second stability criterion we start from the definition of the baryonic mass $M_B$\cite{Mauro}
\begin{equation}
M_B=m_N\int_{0}^{R}\frac{4\pi r^2 n_B(r)}{[1-2Gm(r)/r]^{1/2}}dr,
\end{equation}
where $m_N$ is the neutron mass and $n_B(r)$ is the baryonic density. The regions where $ M/M_B<1 $ are  stable, while those where $M/M_B>1$ correspond  to unstable stellar configurations. Therefore, these regions should also be discarded. But, as shown in the right panel of Fig. \ref{megamma}, all the stellar configurations here obtained are stable with respect to this stability criterion.

\subsection{Comparison of the model with Strange Stars candidates}

In this subsection, we compare the spheroidal static configurations obtained from $\gamma$-equations with some observable data for candidates to be SSs (See Table \ref{candidatas}). As spheroidal SSs have two main radii, for the comparison we defined a mean radius $R_m$, so that the sphere it determines is equal to the surface of the spheroidal star
\begin{equation}
A=2\pi R\left[R+\frac{Z}{\epsilon}\arcsin\epsilon\right],
\end{equation}
where $\epsilon$ is the ellipticity. In this way, the radius $R_m$ could be connected with the surface of the COs and consequently, with its emission properties\cite{Debarati}.

\begin{figure}[h!]
	\centering
	\begin{tabular}{cc}
		\includegraphics[width=0.5\textwidth]{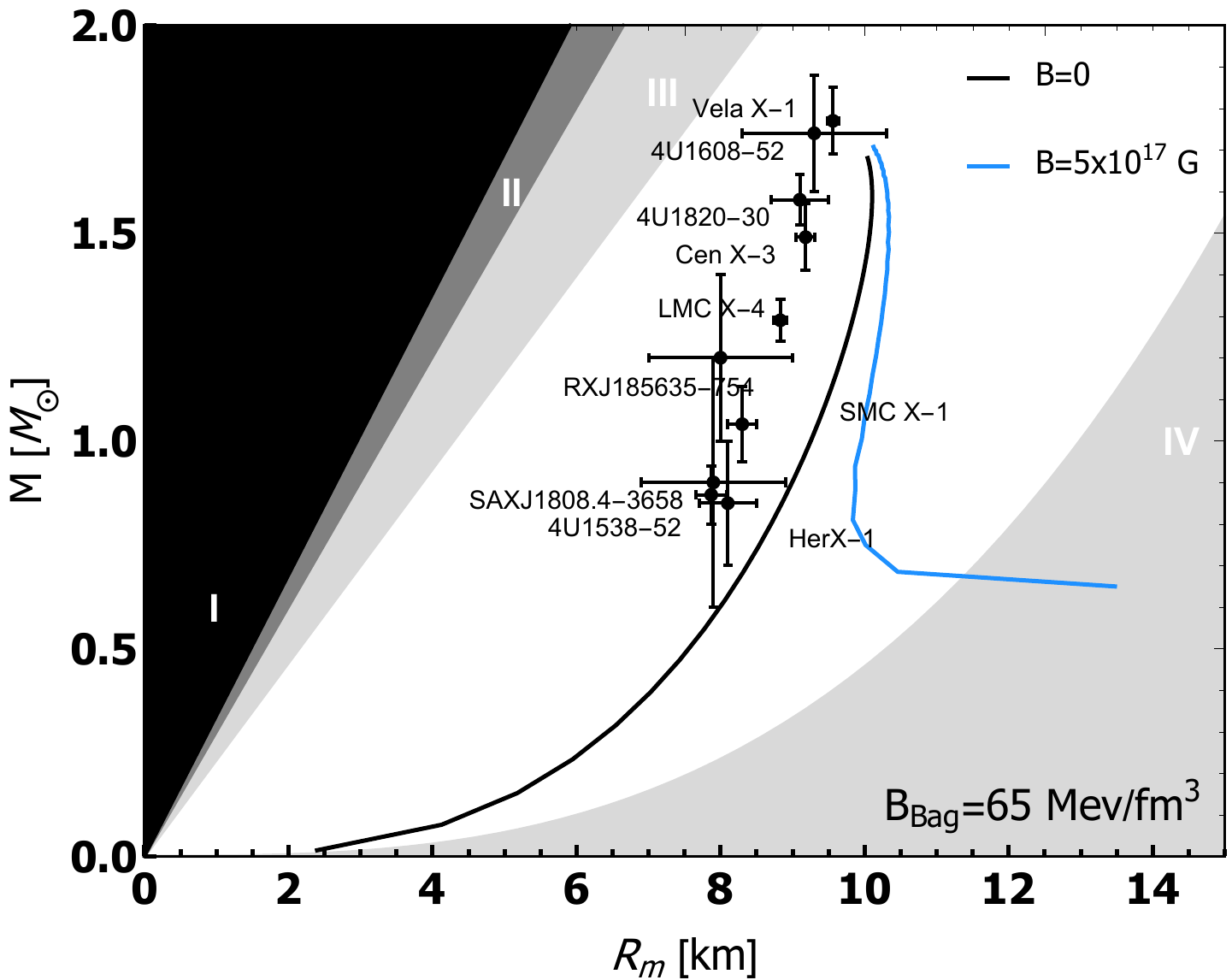}
		& \includegraphics[width=0.5\textwidth]{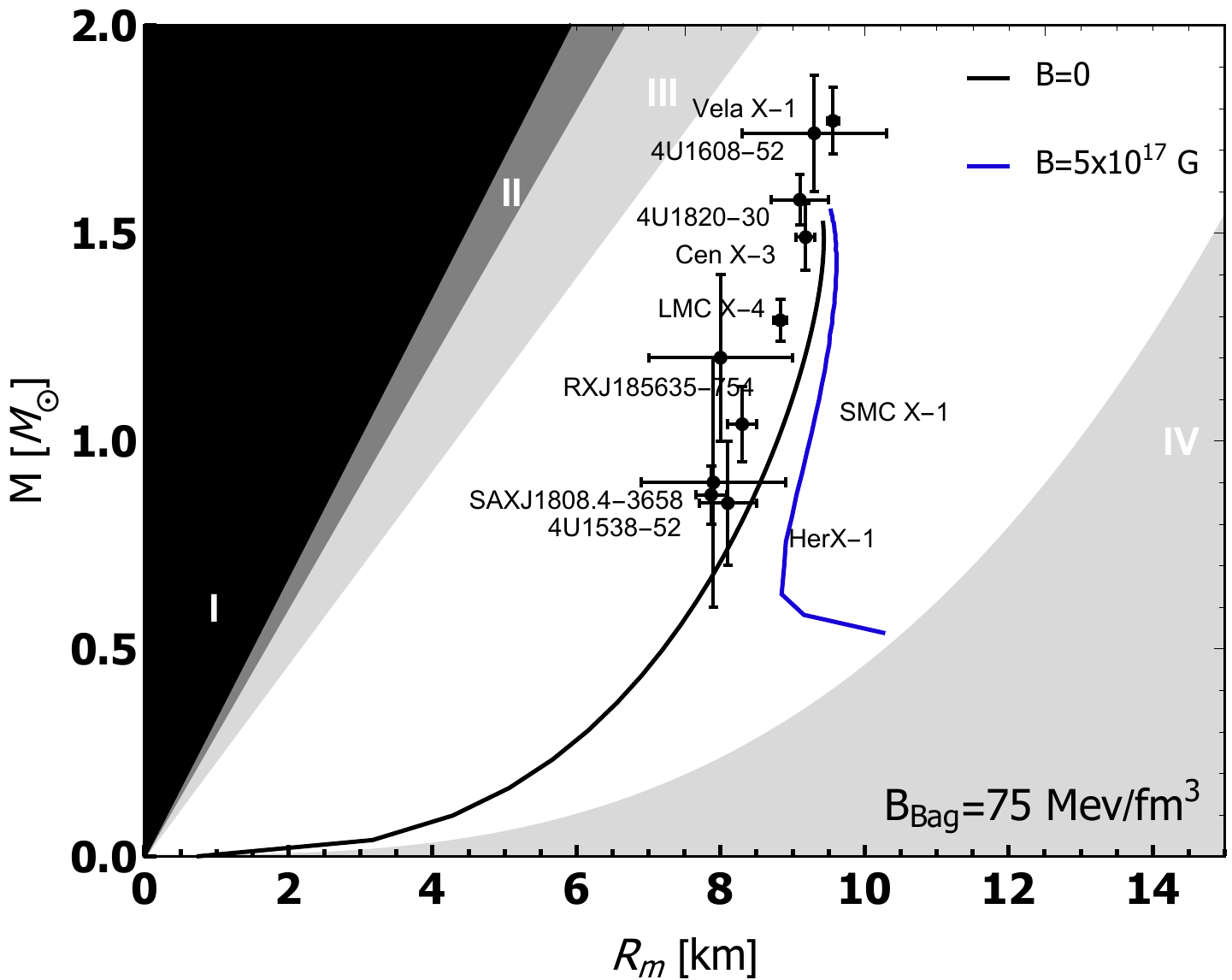}
	\end{tabular}
	\caption{Comparison of observational data for candidates of SSs with the stars obtained from Eqs.~\eqref{EoSSS} and \eqref{gTOV}. The shaded regions correspond to theoretical constraints. Gravitational collapse (I). Requirement of finite pressure inside the star (II). Causality (III). Rotational stability (IV) \cite{Lattimer_prog}. Left panel: $\mathtt{B_{Bag}}=65$~MeV/fm$^3$. Right panel: $\mathtt{B_{Bag}}=75$~MeV/fm$^3$.}
\label{mrcang}
\end{figure}

Fig.\ref{mrcang} shows our theoretical results for the masses and radii together with the observational ones for the candidates. The shaded regions correspond to regions forbidden by theoretical requirements (see the legend of Fig.\ref{mrcang}). The curves at $B=0$ are in the allowed region. In the case of the curves at $B=5\times10^{17}$~G, only the one corresponding to $\mathtt{B_{Bag}}=65$~MeV$/$fm$^3$ has a small  section in the region of rotational instability. In general, our curves are in the region where the candidates are concentrated. In particular, the values of the mass and radius of the stars Vela X-1 and 4U1608-52 fall on the curve $B = 5\times 10^{17}$~G for $\mathtt{B_ {bag}}=65$~MeV$/$fm$^3$. This supports our model of magnetized Strange Stars, meaning it is consistent with the observations.
\section{Magnetized Strange Stars: Redshift and Mass Quadrupole}\label{sec4}

\subsection{Gravitational Redshift}

Gravitational redshift is one of the main effects predicted by the Theory of General Relativity and in turn, constitutes one of its fundamental tests. The frequency of an atomic clock depends on the value of the gravitational potential in the place of its location. So when a photon is observed from a point at a higher gravitational potential, its wavelength is redshifted.

For a spheroidal static CO, the redshift is given by \cite{Zubairi2015}
\begin{equation}
z_{rs}=\frac{1}{\big(1-\frac{2M}{R}\big)^{\gamma/2}}-1,\label{zgamma}
\end{equation}
where $M$ and $R$ are, respectively, the mass and the equatorial radius of the CO. If $\gamma = 1$ the spherical case is recovered.

\begin{figure}[h]
	\centering
	\begin{tabular}{cc}
		\includegraphics[width=0.5\textwidth]{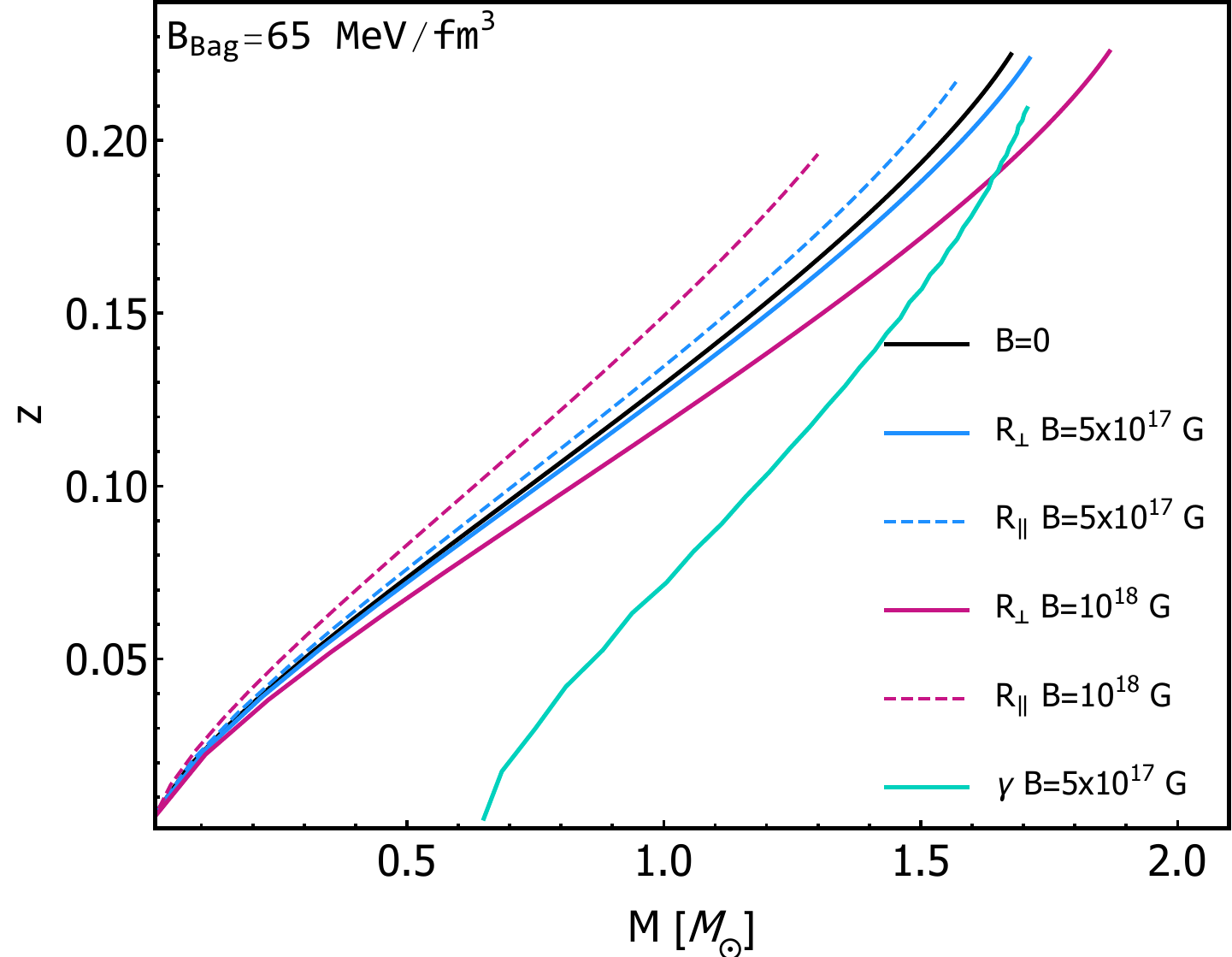}
		& \includegraphics[width=0.5\textwidth]{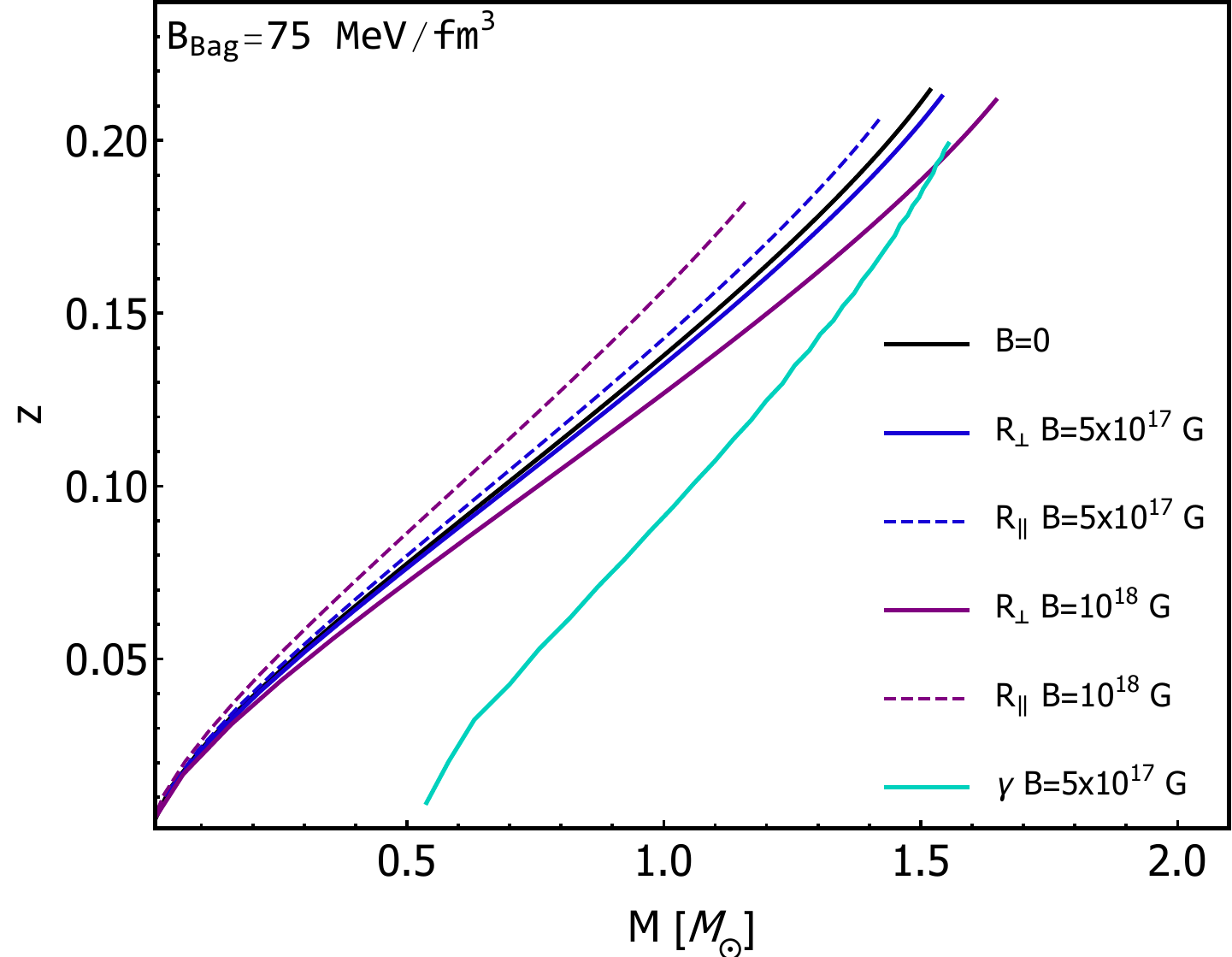}
	\end{tabular}
	\caption{Gravitacional redshift ($z_{rs}$) as a function of the mass ($M$) at $B=[0, 5\times10^{17}, 10^{18}$~G], for TOV and Eqs. $\gamma$. Left panel:  $\mathtt{B_{Bag}}=65$~MeV$/$fm$^3$. Right panel:  $\mathtt{B_{Bag}}=75$~MeV$/$fm$^3$.}
	\label{zmtov}
\end{figure}

The definition of $z_{rs}$ explicitly includes the term $M/R$, so its measurement can narrow these parameters. In addition, $z_{rs}$ has a maximum at the star's maximum point of mass, so it can also be used to rule out EoS that do not lead to observable redshifts and therefore constitutes a benchmark to evaluate the feasibility of NSs models \cite{Camenzind}. Our results for $z_{rs}$ are shown in Fig.~\ref{zmtov} for TOV and  $\gamma$-equations solutions.

In the spherical case, the higher values for $z_{rs}$ are obtained for the pairs ($E, P_{\perp}$). In this case, for a fixed $\mathtt{B_{Bag}}$, increasing the magnetic field increases $z_{rs}$, while changing $\mathtt{B_{Bag}}$ barely affects it. If we compare the results for $B=0$ with the solutions of $\gamma$-equations for $B = 5\times 10^{17}$~G, we see that with these, lower values of $z_{rs}$ are obtained and the curves are far apart. This difference can be very useful to discriminate between models with/without magnetic field  when comparing with observational values.

\subsection{Mass quadrupole moment}

Finally, we calculate the mass quadrupole moment of the spheroidal stars. This magnitude is directly related to the amplitude of the GWs \cite{Magdalena}, since they are only emitted in situations where an asymmetry of mass is generated that gives rise to a quadrupolar moment. Therefore, spherical stars do not have a quadrupole moment and cannot generate GWs. In contrast, magnetized stars, being deformed, have non-zero quadrupole moments that will contribute to their GWs emission. In the framework of our structure equations, the quadrupole moment of the SSs is \cite {Herrera99}
\begin{equation}\label{Mq}
Q=\frac{\gamma^3}{3}M^3(1-\gamma^2).
\end{equation}
where for $\gamma =1$, $Q=0$, as corresponds to the spherical case.

\begin{figure}[h!]
	\centering
	\begin{tabular}{c}
		\includegraphics[width=0.5\textwidth]{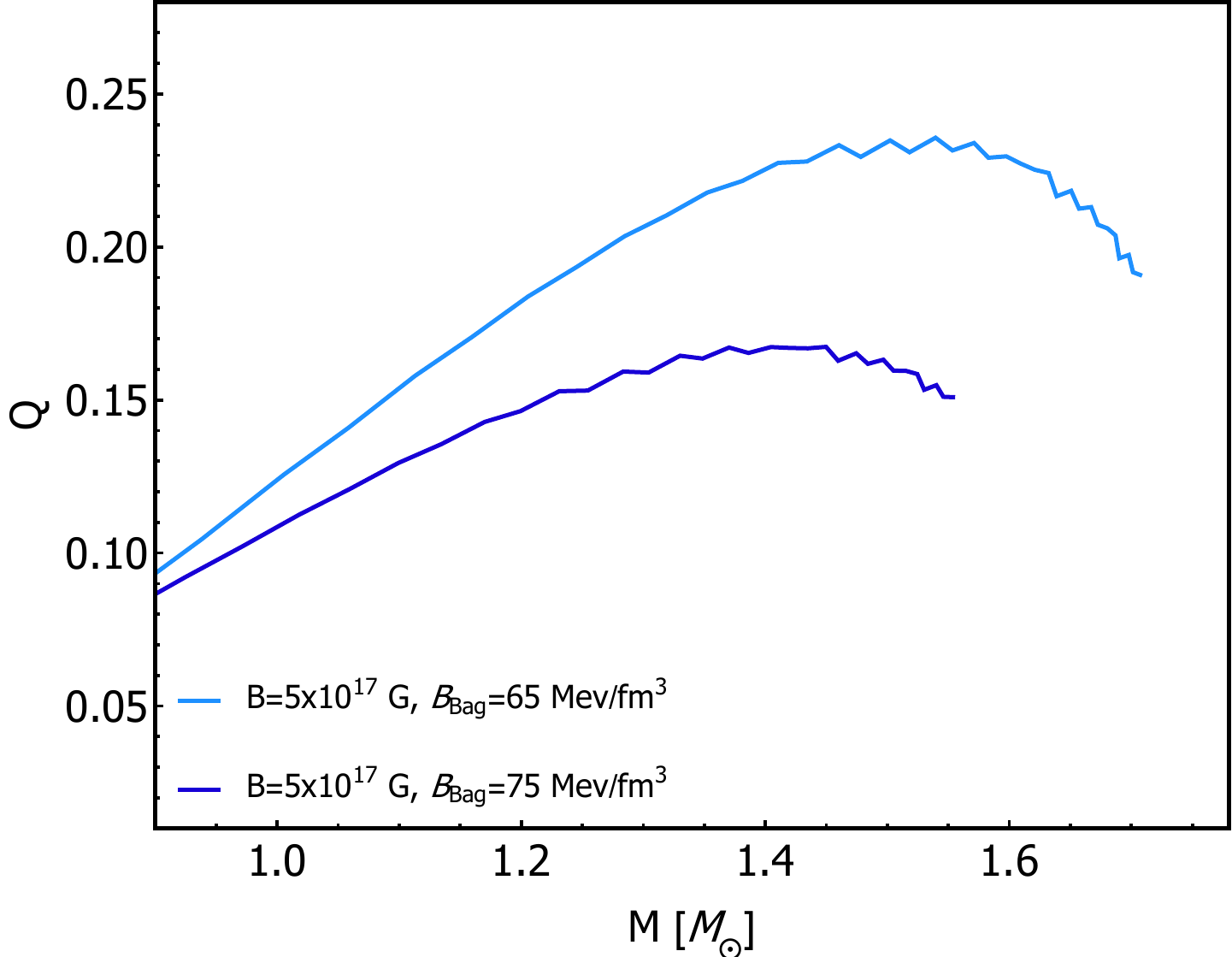}
	\end{tabular}
	\caption{Mass quadrupole moment ($Q$) as a function of the mass ($M$) at $B=5\times10^{17}$~G for $\mathtt{B_{Bag}}=65$~MeV$/$fm$^3$ and $\mathtt{B_{Bag}}=75$~MeV$/$fm$^3$.}
	\label{qmgamma}
\end{figure}

Fig.\ref{qmgamma} shows the SSs quadrupole moment as a function of the star mass. The oscillations in the curve are an effect of the presence of the sum by the Landau levels in the EoS. $Q$ diminishes with $\mathtt{B_{Bag}}$ and its maximum is reached for stars in the region of intermediate masses and deformation. This behavior is due to the simultaneous dependence of $Q$ on $M$ and $\gamma$ and, in particular, is determined by the fact that $\gamma$ depends on the EoS and therefore varies between the stars. This result is different from the one obtained in \cite{Zubairi2015}, where structure equations derived from the $\gamma$ metric were solved by taking $\gamma$ as a free parameter set by hand. In that case, since the mass quadrupole is leaded by the mass, its highest values are reached for the more massive stars. Therefore, connecting $\gamma$ with the physics of the problem has a direct impact on the observables, that again could serve as a way to discriminate between models.


\section{Conclusions}\label{sec5}

This work is an step forward in our studies about magnetized SSs \cite{Felipe2008,Felipe2009JPhG, ARD}. We re-analyzed the stability of SQM under the action of a magnetic field and found that its presence reinforces the Bodmer-Witten's conjecture. To compute the star's observables the EoS have been restricted to the stability region (energies below $930$~MeV). The macroscopic properties of the SSs --mass, radius, deformation, gravitational redshift, mass quadrupole moment-- were calculated using the $\gamma$-equations~\cite{prd} for spheroidal compact objects. The mass--radius curves of the stable configurations obtained are consistent with the observed properties of SSs candidates and comply with the theoretical constraints; this result support the plausibility of our model.

In our model, less massive stars suffer bigger deformations in contrast with the results from TOV solutions for the perpendicular and the parallel pressure independently. This reveals the model--dependency of the results and highlights how important is the construction of even more realistic models. On the other hand, augmenting both, $\mathtt{B_{Bag}}$ or $B$, increases the deformation. However, the effects of changing the bag energy and the magnetic field in the stability with respect to radial oscillations of the solutions of $\gamma$--equations opposes: the stability region increases with $\mathtt{B_{Bag}}$ and decreases with $B$. Nevertheless, none of these parameters influences the stability of the star with respect to the baryon mass criterion.

Another interesting feature of our model is that the redshift and the mass quadrupole moment depend explicitly on the deformation through the EoS --because $\gamma$ appears on their mathematical expressions. As a consequence, the curve of the  gravitational redshift of SSs has remarkable differences with respect to the ones of the spherical case. Besides, the maximum values of the mass quadrupole moment occur for the stars with intermediate values of masses and deformation, suggesting that these are the stars that should produce the most intense GWs emission, instead of those that have the maximum masses or deformation. This result is as interesting as unexpected and deserves a deeper research in a future work.

\section*{Acknowledgments}

The authors have been supported by the grant No. 500.03401 of the PNCB-MES Cuba and by the ICTP Office of External Activities through NT-09.

%

\end{document}